\documentclass[final]{siamltex}

\usepackage{amsmath}
\usepackage{amssymb}
\usepackage{epsfig}
\usepackage{hyperref}

\newcommand{\eqnref}[1]{(\ref{eqn:#1})}
\newcommand{\secref}[1]{Section~\ref{sec:#1}}
\newcommand{\figref}[1]{Figure~\ref{fig:#1}}
\newcommand{\tabref}[1]{Table~\ref{tab:#1}}
\newcommand{\eqnrefs}[1]{(\ref*{eqn:#1})}

\newcommand{\re}[1]{\ensuremath{\mathrm{Re}\left\{#1\right\}}}

\title{Error and symmetry analysis of Misner's algorithm for spherical harmonic
decomposition on a cubic grid}

\author{David R.\ Fiske\thanks{Department of Physics, University of Maryland, 
College Park, MD 20742, and
Laboratory for Gravitational Astrophysics, NASA Goddard Space Flight Center,
Greenbelt, MD 20771, Current Address: Intelligent Systems Division,
Decisive Analytics Corporation, 1235 South Clark Street, Arlington, VA 22202}}

\begin{document}
\maketitle

\begin{abstract}
Computing spherical harmonic decompositions is a ubiquitous technique that
arises in a wide variety of disciplines and a large number of scientific
codes.  Because spherical harmonics are defined by integrals over spheres,
however, one must perform some sort of interpolation in order to compute
them when data is stored on a cubic lattice.
Misner (2004, \textit{Class.\ Quant.\ Grav.}, \textbf{21}, S243)
presented a novel algorithm for computing the spherical harmonic components 
of data represented on a cubic grid, which has been found in
real applications to be both efficient and robust to the presence of
mesh refinement boundaries.  At the same time, however, practical
applications of the algorithm require knowledge of how the truncation
errors of the algorithm depend on the various parameters in the
algorithm.  Based on analytic arguments and experience using the
algorithm in real numerical simulations, I explore these dependencies and
provide a rule of thumb for choosing the parameters based on the
truncation errors of the underlying data.  I also demonstrate that
symmetries in the spherical harmonics themselves allow for an even more
efficient implementation of the algorithm than was suggested by Misner
in his original paper.
\end{abstract}

\section{Introduction}
\label{sec:intro}
Spherical harmonic decomposition is a ubiquitous mathematical technique that
arises in a wide variety of disciplines.  It is no surprise, therefore,
that it also arises in an equally varied range of scientific computing
applications.  It plays a key role, for example, in currently 
active algorithms, for
determining compatible docking configurations of organic
molecules \cite{kovacs:rigid_matching,wriggers:modeling_tricks}, 
modeling atmospheric and oceanic 
flows \cite{evans:atmos_transfer,ocean:circulation,weimer:high-alt},
representing human population density on the surface of the 
earth \cite{tobler:world_pop}, analyzing experimental data 
regarding turbulent magnetohydrodynamic flows \cite{sissan}, 
and analyzing numerical simulations of
gravitational radiation emitted from the collisions of 
black holes \cite{fiske:phd,goddard:teuk}.

In numerical calculations structured on a cubic
grid, however, extracting spherical harmonic components
can be non-trivial since the spherical harmonic components $\Phi_{lm}$
of a function $\Phi$ are defined by integrals over spheres
\begin{equation}
\Phi_{lm}(t,r) 
= \oint \bar{Y}_{lm}(\theta,\phi) \Phi(t,r,\theta,\phi) d\Omega
\label{eqn:sphharm0}
\end{equation}
that have few if any intersections with the cubic lattice.
Misner recently presented a particularly nice
approach to solving this problem, which does not require explicit 
interpolations from the cubic grid to an integration sphere \cite{misner:Ylm}.
The mathematical grounding of the algorithm is laid out completely in
Ref.~\cite{misner:Ylm}, but the original reference does not provide any 
detailed analysis of the numerical errors incurred by the approximation.  In
practical applications, 
such error estimates have been found essential for
choosing the parameters of the algorithm and for understanding the results
of simulations \cite{fiske:phd,goddard:teuk}.
In addition, making use of the symmetries
of the spherical harmonics allows certain simplifications of the
algorithm as applied to generic data and, especially, as applied to data 
with explicit grid symmetries when only the independent portion of the data
is evolved.

It should be noted that although, in my discussion, I will always speak of
spherical harmonics in their usual sense, everything in this paper is also
true for spin-weighted spherical harmonics (see 
Refs.~\cite{Goldberg:swYlm,NP:swYlm} for definitions)
except for \secref{YlmSym}; spin-weighted spherical harmonics do not
generally have the simple symmetries that the (usual) scalar 
spherical harmonics have.
Note also that I follow Misner in treating only the case of uniform grids,
although this method was successfully applied to non-uniform grids
using fixed mesh refinement.  For more details on applications using 
spin-weighted harmonics and run on non-uniform grids, see 
Refs.~\cite{fiske:phd,goddard:teuk}.

The rest of the paper is organized as follows: In \secref{YlmMethod}, I provide
a brief summary of Misner algorithm in order to establish notation for the
following sections.  The error analysis for the algorithm is contained
in \secref{YlmErr}, which leads directly to \secref{YlmParam}, where I
lay out a rule of thumb for choosing the various parameters in the algorithm.
In \secref{YlmSym}, I provide a detailed analysis of the issues 
associated with and simplifications resulting from various symmetries in the
spherical harmonics, and how those symmetries can be exploited for data that
possesses explicit symmetries.

\section{Methodology}
\label{sec:YlmMethod}
In this section I outline in some detail
the steps of the Misner algorithm, although I do not provide a full 
derivation. (See Ref.\ \cite{misner:Ylm} for the derivation.)  
The purpose of this chapter is primarily to establish notation and conventions
for the rest of the paper.

In order to begin, two definitions are need.  First define a radial function
\begin{equation}
R_{n}(r;R,\Delta) = r^{-1}\sqrt{\frac{2n+1}{2\Delta}}
P_{n}\left(\frac{r-R}{\Delta}\right)
\label{eqn:Rn}
\end{equation}
in terms of the usual Legendre polynomials $P_n$.  Here $R$ and $\Delta$ are
parameters that will be associated with the radius at which the spherical
harmonic decomposition is desired and half of the thickness of a shell 
centered on that radius.  From this, define
\begin{equation}
Y_{nlm}(r,\theta,\phi) = R_{n}(r)Y_{lm}(\theta,\phi)
\end{equation}
which form a complete, orthonormal set with respect to the inner product
\begin{equation}
\langle f | g \rangle = \int_{S} \bar{f}(x) g(x) d^{3}x
\label{eqn:continuumIP}
\end{equation}
on the shell
\( S = \{ (r,\theta,\phi) \mid r \in [R-\Delta,R+\Delta] \} \).
Note also that, because the functions $R_n$ form a complete set on the shell,
\begin{equation}
\Phi_{lm}(t,R) = \int \left[ 
	\sum_{n=0}^{\infty} R_{n}(R;R,\Delta)R_{n}(r;R,\Delta) \right]
	\bar{Y}_{lm}(\theta,\phi) \Phi(r,\theta,\phi) d^{3}x
\label{eqn:PhiLMexact}
\end{equation}
and that the term in brackets 
\begin{equation}
\sum_{n=0}^{\infty} R_{n}(R;R,\Delta)R_{n}(r;R,\Delta)
	= r^{-2}\delta(R-r)
\label{eqn:delta}
\end{equation}
is a delta function.\footnote{The Dirac delta function is defined by
the property that, for any function $f$, the integral
\(\int_{a}^{b} f(y) \delta(x-y) dy\) is $f(x)$ when
\(x\in(a,b)\) and is 0 otherwise.} 
(Compare \eqnref{PhiLMexact} to \eqnref{sphharm0}.)

On a finite grid $\Gamma$, the inner product \eqnref{continuumIP}
will have the form
\begin{equation}
\langle f | g \rangle = \sum_{x\in\Gamma} \bar{f}(x) g(x) w_{x}
\label{eqn:numip}
\end{equation}
where each point has some weight $w_{x}$.  This weight was given the
form
\begin{equation}
w_{x} = \left\{ 
\begin{array}{ll}
0   & |r-R| > \Delta + h/2 \\
h^3 & |r-R| < \Delta - h/2 \\
(\Delta +h/2 - |R-r|)h^2 & \mbox{otherwise}
\end{array}\right.
\label{eqn:MisnerWeights}
\end{equation}
by Misner, where $h$ is the grid spacing.
Only cases with $\Delta > h/2$ are considered.
This means, roughly, that points entirely within the 
shell $S$ are weighted by their finite volume on the numerical grid,
points entirely outside of the shell $S$ have zero weight, and points near
the boundary are weighted according to the fraction of their volume 
inside $S$.\footnote{Actually the boundary points are weighted by the fraction
of their volume that would be inside $S$ if the point were on a coordinate
axis.  See Ref.\ \cite{misner:Ylm} for more details on the definition of the 
weights.}

With the numerical inner product \eqnref{numip}, and letting capital
Roman letters \(A = (nlm)\) represent index groups, the $Y_{A}$ are no
longer orthonormal.  Their inner product
\begin{equation}
\langle Y_A | Y_B \rangle = G_{AB} = \bar{G}_{BA}
\label{eqn:YlmMetric}
\end{equation}
forms a metric for functions on the shell.  Although a priori this matrix
appears to be complex valued, I will show in \secref{YlmSym} that it is 
actually real-symmetric and sparse.  For now it suffices to follow Misner
in denoting it as generically Hermitian.
The inverse to this metric
$G^{AB}$ can be used to raise indices on functions defined on the sphere.

Making use of this new metric, and with some further analysis, the
approximation for the spherical harmonic coefficients
\begin{equation}
\Phi_{lm}(t,R) = \sum_{x\in\Gamma}\bar{R}_{lm}(x;R)w_{x}\Phi(t,x)
\label{eqn:PhiLMnum}
\end{equation}
follows with
\begin{equation}
R_{lm}(x;R) = \sum_{n=0}^{N} \bar{R}_{n}(R)Y^{nlm}(x)
\label{eqn:Rlm}
\end{equation}
in terms of $Y^{A} = G^{BA}Y_{B}$, not $Y_{A}$.

\section{Error Analysis}
\label{sec:YlmErr}
In Ref.~\cite{misner:Ylm}, Misner provides an algorithm for computing
spherical harmonic components on a cubic lattice, but he does not provide
any analysis of the errors involved.  In a practical application of the
method, such error analysis, in particular an analysis of the convergence
order in grid spacing, is very valuable.  For this reason, I wish to 
examine the issue more closely here.

To begin the analysis, note that there are three parameters in the
algorithm, the grid spacing $h$, the width of the shell $\Delta$, and
the number of terms in the sum over basis polynomials $N$.  
(Compare \eqnref{PhiLMexact} to \eqnref{PhiLMnum} and \eqnref{Rlm}.)
In the limit that $h \rightarrow 0$ and $N \rightarrow \infty$, the numerical
algorithm goes to the continuum theory.  Note that, when $h$ and $N$
go to their limits, \emph{any} value for
$\Delta$ is allowed, since \eqnref{PhiLMexact} is an exact, continuum
expression.  For a finite value of $N$, however, the quality of the 
approximation in the radial direction is a function of both $N$ and $\Delta$.
Moreover, tying the value of $\Delta$ to the grid spacing $h$, though not
necessary from fundamental considerations, does in fact have advantages in
terms of convergence behavior that will become clear below.

I first consider the behavior of the algorithm as a function of $\Delta$
for fixed values of $N$. Define
\begin{equation}
d(x;N,R,\Delta) = \sum_{n=0}^{N} \frac{2n+1}{2\Delta}
P_{n}\left(\frac{x-R}{\Delta}\right) P_{n}(0)
\label{eqn:d}
\end{equation}
which is closely related to the delta function \eqnref{delta} in the limit
$N \rightarrow \infty$.  For finite $N$, this should approximate the
delta function to some order of accuracy.
To quantify this, consider the approximation
\begin{equation}
f(0) \approx I_{N}[f] \equiv \int_{-\Delta}^{\Delta} d(x;N,0,\Delta)f(x)dx
\label{eqn:dconv1}
\end{equation}
for a suitably smooth test function $f$.  In the limit that 
$N \rightarrow \infty$, this is exact.  For any finite $N$, this expression
can be analyzed by Taylor expanding the test function around the point
$x=0$.  This gives, after rearranging terms and noting that all of the odd
terms vanish by symmetry,
\begin{equation}
I_{N}[f] = \sum_{k=0}^{\infty} \frac{c_{N,2k}}{(2k)!} f^{(2k)}(0)
\end{equation}
where
\begin{equation}
c_{N,k}(\Delta) = \int_{-\Delta}^{\Delta} x^{k} d(x;N,0,\Delta) dx
\end{equation}
are the coefficients of the expansion.  In order to have a high order method,
I need $c_{N,0}=1$, and the coefficients corresponding to the
next few values of $k$ to vanish.  \tabref{cnk} shows the first few
\begin{table}
\centering
\begin{tabular}{|r||r|r|r|r|r|r|r|} \hline
$_{N}$\verb.\.$^{k}$ & 0 & 2 & 4 & 6 & 8 & 10 & 12 \\ \hline \hline
0 & 1 & 1/3 & 1/5 & 1/7 & 1/9 & 1/11 & 1/13 \\ \hline
2 & 1 & 0 & -3/35 & -2/21 & -1/11 & -12/143 & -1/13 \\ \hline
4 & 1 & 0 & 0 & 5/231 & 5/143 & 6/143 & 10/221 \\ \hline
6 & 1 & 0 & 0 & 0 & -7/1287 & -23/2431 & -70/4199 \\ \hline
8 & 1 & 0 & 0 & 0 & 0 & 63/46189 & 15/4199 \\ \hline
10 & 1 & 0 & 0 & 0 & 0 & 0 & -33/96577 \\ \hline
\end{tabular}
\caption{The first few non-trivial values of $c_{N,k}(1)$, which are the
coefficients of the Taylor expansion of a function integrated against
$d(x;N,0,1)$.  (In general, $c_{N,k}(\Delta) = c_{N,k}(1)\Delta^{k}$.)
The values of $k$ run across and the values of $N$ run down.
The fact that the first coefficient is always 1, and
that, by increasing $N$, more of the sub-leading coefficients are 0 indicates
that increasing $N$ increases the order of convergence of the Misner
algorithm (provided that $\Delta \propto h$).}
\label{tab:cnk}
\end{table}
coefficients \(c_{N,k}(1) = \Delta^{-k}c_{N,k}(\Delta)\)
for even $N$ in [0,10].  It is clear that each time
$N$ is increased, one more of the sub-leading coefficients vanishes.  This
means that there are error terms proportional to $\Delta$, and that the leading
order term arises at $\mathcal{O}(\Delta^{N+2})$.  
In order to prevent this term,
at high resolution, from dominating over discretization errors scaling like
powers of the grid spacing, one should choose $\Delta \propto h$.

Additional error terms proportional the grid spacing $h$ depend
primarily on the order of accuracy in the volume integral when using the
weights defined by \eqnref{MisnerWeights}.  For a finite volume, this weighing
scheme provides an approximation that scales as $\mathcal{O}(h)$, but, for
a region that is also scaling with $h$, the resulting integral scales as
$\mathcal{O}(h^2)$.  This provides additional motivation for choosing
\(\Delta \propto h\) since most applications will require at least second order
accuracy.

For higher than second order accuracy, a new scheme for computing
the volume integrals is needed, but the rest of the analysis here holds true.
Given such a scheme, the analysis here shows how to choose the remaining
parameters to ensure that numerical errors scale like any desired power of the
grid spacing.

\section{Choosing the parameters}
\label{sec:YlmParam}
In practice, the grid spacing parameter $h$ is usually chosen to resolve the
sources without exceeding the physical limits of the computer.  I would
not expect, in general, that the grid spacing would be chosen based on the
needs of this algorithm.  For that reason, let me assume now that
$h$ is chosen, and discuss how to choose the remaining parameters
$N$ and $\Delta$.  In this section I will discuss some of the
theoretical issues that should be considered when choosing the parameters,
leading to a rule of thumb that is valid based on this analysis and my
experience with the algorithm.

The error analysis of \secref{YlmErr} implies that for fixed $\Delta$,
increasing the value of $N$ decreases the error term.
It also implies that for fixed $N$, increasing the value of $\Delta$
increases the error term.  This suggests taking $\Delta$ as small as 
possible, and $N$ as large as possible to make the error term as small
as possible.
This must be balanced, however, against practical limitations.  Certainly
the shell thickness $\Delta$ needs to be large enough so that there are
some grid points within the shell, otherwise the whole procedure is
undefined.  For fixed $N$, a stronger restriction requires that the
Legendre polynomial $P_{N}$ can be resolved over the shell.  Without this
condition, there would seem to be no benefit to taking higher values of $N$.
Getting higher accuracy in practice requires finding a proper balance between
choosing $\Delta$ small and $N$ large.

In making this balance, however, one must keep in mind that the error in
the method is partially determined by the weighting scheme
\eqnref{MisnerWeights}, which is only second order accurate in the grid
spacing.  I am, in addition, going to choose $\Delta \propto h$ for 
reasons described above.  This already suggests that taking $N$ larger than
two is pointless, since choosing $N=2$ already makes the piece of the error
that is proportional to $\Delta$ scale like $\mathcal{O}(\Delta^4)$ (cf.\
\tabref{cnk}), meaning that it will be an error term of sub-leading order
in grid spacing.  But once this term is of sub-leading order, it is much
less important how large I choose $\Delta$, provided that I still choose
it proportional to the grid spacing.  I therefore adopt the following
\begin{quote}
\textbf{Rule of Thumb}: Choose $N$ just large enough to ensure that
the error term proportional to $\Delta$ is an error term of sub-leading
order in grid spacing.  Choose $\Delta$ just large enough to safely
resolve $P_{N}$ on the shell.
\end{quote}
With this rule of thumb, and the second order accurate weighting
scheme \eqnref{MisnerWeights}, I found the choices $N=2$ and
$\Delta = 3h/4$ completely satisfactory for a second order accurate code.  
Note that this corresponds to
Misner's choice of $\Delta$ in Ref.\ \cite{misner:Ylm}.  With
$N=2$ I found that larger values of $\Delta$ are also acceptable.
Numerical results justifying these estimates appear in Ref.~\cite{fiske:phd}.

\begin{figure}
\begin{center}
\epsfig{file=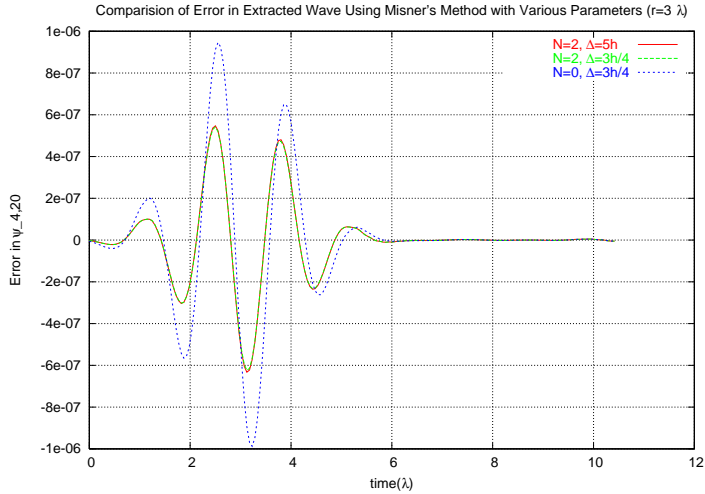,scale=0.75}
\end{center}
\caption{A comparison of numerical results as a function of parameters.
The lines show \emph{errors} in a numerical solution as compared to
an analytic solution for a sample problem.  When $N=2$ the errors are
smaller than when $N=0$, consistent with the fact that the errors
scale like the square of grid spacing for $N=2$ and only like the grid
spacing for $N=0$.  For fixed $N$, the size of $\Delta$ is fairly unimportant.}
\label{fig:paramCompare}
\end{figure}

This is illustrated in \figref{paramCompare},
which shows the errors in a test function as compared to an analytic solution.
(The physical problem is the propagation of a linear gravitational
wave through a mesh refined grid, as described in 
Refs.~\cite{fiske:phd,goddard:teuk}.)  It is clear from the figure that
choosing $N$ large enough makes a dramatic impact on the truncation error,
whereas the exact value of $\Delta$ for fixed $N$ is less important.
In this test case, the underlying simulation was only second order accurate,
so there is no benefit from choosing $N>2$.

\section{Symmetry issues}
\label{sec:YlmSym}
There are two points of interest related to this method of spherical
harmonic decomposition and symmetries.  The first was mentioned briefly
in \secref{YlmMethod}, namely that symmetry causes the metric
$G_{AB}$ to be real and sparse.  The second deals with implementing the method
for grids in which explicit symmetries are enforced on grid functions
in order to reduce the computational load of the simulation.  In these cases,
in which data is not evolved over a whole extraction sphere, additional
analysis is required to demonstrate that the method is well defined and to
understand how to most efficiently implement it.  The primary result on
this second topic is that the adjoint harmonics $Y^A$ of \eqnref{Rlm} have the
same symmetries under reflection as the original spherical harmonics $Y_A$.

Consider first the implications of symmetry on the metric $G_{AB}$.
The symmetries of the spherical harmonics, summarized in \tabref{YlmReflect}, 
\begin{table}
\centering
\begin{tabular}{|r|r|r|l|c|} \hline
\textbf{Planes} & $\theta$ & $\phi$ & \textbf{Sign} & \textbf{Conjugate} \\ 
	\hline \hline
None & $\theta$ & $\phi$ & $+1$ & no \\ \hline
$x$ & $\theta$ & $\pi-\phi$ & $(-1)^m$ & yes \\ \hline
$y$ & $\theta$ & $-\phi$ & $+1$ & yes \\ \hline
$z$ & $\pi-\theta$ & $\phi$ & $(-1)^{l+m}$ & no \\ \hline
$xy$ & $\theta$ & $\pi+\phi$ & $(-1)^m$ & no \\ \hline
$xz$ & $\pi-\theta$ & $\pi-\phi$ & $(-1)^{l}$ & yes \\ \hline
$yz$ & $\pi-\theta$ & $-\phi$ & $(-1)^{l+m}$ & yes \\ \hline
$xyz$ & $\pi-\theta$ & $\pi+\phi$ & $(-1)^{l}$ & no \\ \hline
\end{tabular}
\caption{The table shows how the arguments of spherical harmonics transform
under reflections through various Cartesian planes.  The first column indicates
which coordinates have their signs inverted, while the second and 
third columns give the new angular arguments to the spherical harmonic
$Y_{lm}$.  Alternatively, the fourth and fifth column show, respectively,
the overall sign in front of and whether or not to complex conjugate the
given spherical harmonic with the original angular arguments.  The second row,
for example, says that 
$Y_{lm}(-x,y,z) = Y_{lm}(\theta,\pi-\phi) = (-1)^m \bar{Y}_{lm}(\theta,\phi)$,
where $(\theta,\phi)$ are the angular coordinates of the point $(x,y,z)$.
Note that this table differs slightly from that in Ref.~\cite{fiske:phd}.
The table here is correct.}
\label{tab:YlmReflect}
\end{table}
cause the imaginary part of all terms in the integral 
\eqnref{YlmMetric} to cancel in pairs of points on the sphere related by 
reflections through coordinate planes.
The reason is that each of the four signs ($+1$, $(-1)^m$, $(-1)^l$, and
$(-1)^{l+m}$) appears twice in \tabref{YlmReflect}, 
once for a term that is complex
conjugated and once for a term that is not.  
The matrix is also sparse.  By similar reasoning, for certain values
of $l$ and $m$, the terms in the integral \eqnref{YlmMetric} can cancel
in sets of four.  Both of these facts can be seen at once through a simple
calculation.  The idea is to break the integral into parts using the
second and third columns of \tabref{YlmReflect}, and then to simplify
using the last two columns.  Considering first just the symmetries under
reflection through the $xy$-plane and recalling the definition 
\eqnref{YlmMetric},
\begin{subequations}
\begin{eqnarray}
G_{AB} & = &
\oint\bar{Y}_{l_{1}m_{1}}(\theta,\phi)Y_{l_{2}m_{2}}(\theta,\phi)d\Omega \\
 & = & \int_{0}^{2\pi} \int_{0}^{\pi/2}
	\bar{Y}_{l_{1}m_{1}}(\theta,\phi)Y_{l_{2}m_{2}}(\theta,\phi) 
	d\Omega \nonumber \\
 & & \mbox{} + \int_{0}^{2\pi} \int_{0}^{\pi/2} 
	\bar{Y}_{l_{1}m_{1}}(\pi-\theta,\phi)
	Y_{l_{2}m_{2}}(\pi-\theta,\phi) d\Omega \\
 & = & [1+(-1)^{l_{1}+l_{2}}] 
	\int_{0}^{2\pi}\int_{0}^{\pi/2} 
	\bar{Y}_{l_{1}m_{1}}(\theta,\phi)Y_{l_{2}m_{2}}(\theta,\phi)
	d\Omega .
\end{eqnarray}
\end{subequations}
(I have suppressed the radial functions since they play no role here.)
Repeating the procedure for reflections through the $xz$- and $yz$-planes
shows that
\begin{equation}
G_{AB} = 2 \sigma_{m_{1}+m_{2},l_{1}+l_{2}}
\int_{0}^{\pi/2}\int_{0}^{\pi/2} 
\re{\bar{Y}_{l_{1}m_{1}}(\theta,\phi)Y_{l_{2}m_{2}}(\theta,\phi)}d\Omega
\end{equation}
where
\begin{equation}
\sigma_{m_{1}+m_{2},l_{1}+l_{2}} 
\equiv 1 + (-1)^{m_1 + m_2} + (-1)^{l_1 + l_2} + (-1)^{m_1 + m_2 + l_1 + l_2}.
\end{equation}
This proves that the matrix is real.  In addition, the matrix element
is zero by symmetry whenever
\begin{equation}
\sigma_{m_{1}+m_{2},l_{1}+l_{2}} = 0
\label{eqn:YlmParity}
\end{equation}
which is true for 56 of the 81 matrix elements that exist when considering
a fixed value of $n$ and all values of $l$ and $m$ for $l \leq 2$.
Of the remaining 25 matrix elements, 9, of course, are the diagonal elements
that go to unity in the continuum limit.
The exact  break-down of which such elements must be zero by 
symmetry is summarized
in \tabref{YlmParity}.
\begin{table}
\centering
\begin{tabular}{|c|c|c|c|} \hline
$m_1 + m_2$ & $l_1 + l_2$ & Number & Satisfies \eqnref{YlmParity} \\ 
	\hline \hline
even & even & 25 & no \\ \hline
even & odd & 16 & yes \\ \hline
odd & even & 20 & yes \\ \hline
odd & odd & 20 & yes \\ \hline
\end{tabular}
\caption{The table summarizes which entries of $G_{AB}$ identically vanish
because of the symmetries of the spherical harmonics under reflections through
coordinate planes for all values of $l$ and $m$ with $l \leq 2$. 
(This is governed by equation \eqnrefs{YlmParity}.)  Of the 81 possible matrix
elements, only 25 have non-trivial values.}
\label{tab:YlmParity}
\end{table}
Knowing that the matrix is real-symmetric and sparse allows for a more
efficient implementation of the algorithm in general.  It is also
extremely useful in analyzing the algorithm in the context of the
second topic of this section, explicit grid symmetries.

When evolving initial data with known symmetries, it is very common to
evolve only that part of the data that is unique.  In such cases, an
appropriate symmetry boundary condition is applied at some edges of the
grid.  This is, however, inconvenient for wave extraction since computing
spherical harmonic components (by any method) requires integrating over the
full sphere.  If data with octant symmetry, for example, is evolved only
in a single octant, it is neither sufficient to apply the decomposition
algorithm in that one octant nor to multiply the result of a single octant
by 8 since the symmetry may forbid some modes as well as repeating them.

In principle the problem appears to be even more difficult for this
particular decomposition method.  Although the spherical harmonics
have well defined symmetries under reflections, as summarized in
\tabref{YlmReflect}, it is the adjoint harmonics that appear in
\eqnref{Rlm}.  The adjoint harmonics, however, are constructed by
contracting $G^{AB}$ with the usual spherical harmonics, and this appears
to mix different values of $l$ and $m$.  While this mixing does occur,
the the matrix $G_{AB}$ is sparse in just the right way to ensure that
the adjoint harmonics have the same symmetries at the usual spherical
harmonics.

A particular choice of the mapping
\((n,l,m) \mapsto A\) makes this easiest to see.  Specifically, considering
all values of $l$ and $m$ with $l \leq 2$, there is a basis in which
$G_{AB}$ takes block diagonal form
\begin{equation}
(G_{AB}) = \left( \begin{array}{cccc}
\Xi_1 & & & \\
 & \Xi_2 & & \\
 & & \Xi_3 & \\
 & & & \Xi_4
\end{array} \right)
\end{equation}
with all unwritten entries identically zero by symmetry.
In this expression $\Xi_1$ is a $4N \times 4N$ matrix over the basis functions
\( B_1 = \{Y_{n00}, Y_{n2,-2}, Y_{n20}, Y_{n22} \} \); 
$\Xi_2$ is an $N \times N$ matrix over the basis functions 
\( B_2 = \{ Y_{n10} \} \);
$\Xi_3$ is a $2N \times 2N$ matrix over the basis functions
\( B_3 = \{ Y_{n1,-1}, Y_{n11} \} \); 
and $\Xi_4$ is a $2N \times 2N$ matrix over the basis functions 
\( B_4 = \{ Y_{n2,-1}, Y_{n21} \}\).  In this basis the matrix is 
block diagonal, so the inverse matrix
\begin{equation}
(G^{AB}) = \left( \begin{array}{cccc}
\Xi_{1}^{-1} & & & \\
 & \Xi_{2}^{-1} & & \\
 & & \Xi_{3}^{-1} & \\
 & & & \Xi_{4}^{-1}
\end{array} \right)
\end{equation}
is also block diagonal and \emph{the different basis sectors do not mix}.
This last point is key.
It implies that any particular adjoint harmonic
is a linear combination of spherical harmonics from a single set $B_{k}$
\begin{equation}
Y^{nlm} = \sum_{Y_{n'l'm'}\in B_{k}} (\Xi_{k})^{(nlm)(n'l'm')} Y_{n'l'm'}
\end{equation}
where $k$ is the index such that \(Y_{nlm} \in B_{k}\).
Because, in each set $B_k$, the parity of $l$ and the parity of $m$ is the
same on each $Y_{nlm} \in B_k$, and because, in light of \tabref{YlmReflect},
it is the parity of $l$ and $m$ that determines the symmetries of
$Y_{nlm}$ under reflections through planes,
every spherical harmonic in $B_{k}$ for any fixed $k$ has the same
symmetries under reflections as any other spherical harmonic in $B_k$.
This implies that the adjoint harmonics also share this symmetry under
reflection.

\section{Discussion}
\label{sec:discussion}
In this paper I have provided detailed error estimates of the Misner algorithm
for computing spherical harmonic components of data represented on a cubic
grid.  This analysis allows one, in principle,
to chose the parameters of the algorithm
such that its numerical errors scale any desired power of the grid spacing.
The only limitation of this in practice is finding a scheme for approximating
volume integrals on a shell of sufficiently high accuracy. (Misner's original
choice allows for a second order accurate result.)

In addition, analysis of the symmetry properties of the spherical harmonics
provides two insights: First, the number of operations required to initialize
the data structures used to compute spherical harmonic components can be
reduced by computing only those elements of $G_{AB}$ that are not forbidden
by symmetries, and, second, that the adjoint harmonics used by the algorithm
have the same symmetries under reflections as the usual spherical harmonics.
This second fact allows the method to be efficiently used on data with explicit
grid symmetries when only the independent portion of that data is evolved.

\section{Acknowledgments}
I am grateful to Charles Misner, Richard Matzner, and John Baker for 
helpful discussions and suggestions.  This work
was supported by NASA Space Sciences grant ATP02-0043-0056.  Portions of
the text were written while the author held a National Research Council
Post-doctoral Fellowship at NASA Goddard Space Flight Center.

\bibliographystyle{siam}
\bibliography{gr,spherical_harmonic}

\begin{thebibliography}{10}

\bibitem{evans:atmos_transfer}
{\sc K.~F. Evans}, {\em The spherical harmonics discrete ordinate method for
  three-dimensional atmospheric radiative transfer}, J. Atmos. Sci, 55 (1998),
  p.~429.

\bibitem{fiske:phd}
{\sc David~Robert Fiske}, {\em Numerical studies of constraints and
  gravitational wave extraction in general relativity}, PhD thesis, University
  of Maryland, College Park, 2004.
\newblock http://hdl.handle.net/1903/1805.

\bibitem{goddard:teuk}
{\sc David~R. Fiske, John~G. Baker, James~R. van Meter, Dae-Il Choi, and
  Joan~M. Centrella}, {\em Wave zone extraction of gravitational radiation in
  three-dimensional numerical relativity}, Phys. Rev. D, 71 (2005), p.~104036.

\bibitem{Goldberg:swYlm}
{\sc J.~N. Goldberg, A.~J. MacFarlane, E.~T. Newman, F.~Rohrlich, and E.~C.~G.
  Sudarshan}, {\em Spin-$s$ spherical harmonics and $\eth$}, J. Math. Phys., 8
  (1967), p.~2155.

\bibitem{kovacs:rigid_matching}
{\sc Julio~A. Kovacs, Pablo Chac{\'o}n, Yao Cong, Essam Metwally, and Willy
  Wriggers}, {\em Fast rotational matching of rigid bodies by fast {Fourier}
  transform acceleration of five degrees of freedom}, Acta Cryst. D, 59 (2003),
  p.~1371.

\bibitem{misner:Ylm}
{\sc Charles~W. Misner}, {\em Spherical harmonic decomposition on a cubic
  grid}, Class. Quant. Grav., 21 (2004), p.~S243.

\bibitem{NP:swYlm}
{\sc E.~T. Newman and R.~Penrose}, {\em Note on the {B}ondi-{M}etzner-{S}achs
  group}, J. Math. Phys., 7 (1966), p.~863.

\bibitem{sissan}
{\sc Daniel~R. Sissan, Nicol{\'a}s Mujica, W.~Andrew Tillotson, Yi-Min Huang,
  William Dorland, Adil~B. Hassam, Thomas~M. Antonsen, and Daniel~P. Lathrop},
  {\em Experimental observation and characterization of the magnetorotational
  instability}, Phys. Rev. Lett., 93 (2004), p.~114502.

\bibitem{ocean:circulation}
{\sc D.~Stammer, R.~Tokmakian, A.~Semtner, and C.~Wunach}, {\em How well does a
  1/4 degree global circulation model simulate large-scale oceanic
  observations?}, J. Geophys. Res., 101 (1996), p.~25779.

\bibitem{tobler:world_pop}
{\sc Waldo Tobler}, {\em Preliminary representation of world population by
  spherical harmonics}, Proc. Natl. Acad. Sci. USA, 89 (1992), p.~6262.

\bibitem{weimer:high-alt}
{\sc D.~R. Weimer}, {\em Models of high-altitude electric potentials derived
  with a least error fit of spherical harmonic coefficients}, J. Geophys. Res.,
   (1995), p.~19595.

\bibitem{wriggers:modeling_tricks}
{\sc Willy Wriggers and Pablo Chac{\'o}n}, {\em Modeling tricks and fitting
  techniques for multiresolution structures}, Structure, 9 (2001), p.~779.

\end{thebibliography}
\end{document}